\documentclass[preprints,article,accept,moreauthors,pdftex,10pt,a4paper]{mdpi} 

\firstpage{1} 
\pubvolume{xx}
\issuenum{1}
\articlenumber{1}
\pubyear{2018}
\copyrightyear{2018}
\externaleditor{Academic Editor: name}
\history{Received: date; Accepted: date; Published: date}

\Title{Critical Evaluation of Organic Thin-Film Transistor Models}

\Author{Markus Krammer $^{1}$, James W. Borchert $^{2}$, Andreas Petritz $^{3}$, Esther Karner-Petritz $^{3}$, Gerburg Schider $^{3}$, Barbara Stadlober $^{3}$, Hagen Klauk $^{2}$ and Karin Zojer $^{1}$*}

\AuthorNames{Markus Krammer, James W. Borchert, Andreas Petritz, Esther Karner-Petritz, Gerburg Schider, Barbara Stadlober, Hagen Klauk and Karin Zojer}

\address{%
$^{1}$ \quad Institute of Solid State Physics, NAWI Graz, Graz University of Technology, Petersgasse 16, 8010 Graz, Austria; karin.zojer@tugraz.at\\
$^{2}$ \quad Max Planck Institute for Solid State Research, Heisenbergstr. 1, 70569 Stuttgart, Germany; H.Klauk@fkf.mpg.de\\
$^{3}$ \quad Joanneum Research Materials, Institute for Surface Technologies and Photonics, Franz-Pichler-Straße 30, 8160 Weiz, Austria; barbara.stadlober@joanneum.at}

\corres{Correspondence: karin.zojer@tugraz.at; Tel.: +43-316-873-8974}

\abstract{Thin-film transistors (TFTs) represent a wide-spread tool to determine the charge-carrier mobility of materials.
Mobilities and further transistor parameters like contact resistances are commonly extracted from the electrical characteristics.
However, the trust in such extracted parameters is limited, because their values depend on the extraction technique and on the underlying transistor model.
We propose a technique to establish whether a chosen model is adequate to represent the transistor operation.
This two-step technique analyzes the electrical measurements of a series of TFTs with different channel lengths.
The first step extracts the parameters for each individual transistor by fitting the full output and transfer characteristics to the transistor model. 
The second step checks whether the channel-length dependence of the extracted parameters is consistent with the model.
We demonstrate the merit of the technique for distinct sets of organic TFTs that differ in the semiconductor, the contacts, and the geometry.
Independent of the transistor set, our technique consistently reveals that state-of-the-art transistor models fail to reproduce the correct channel-length dependence.
Our technique suggests that contemporary transistor models require improvements in terms of charge-carrier-density dependence of the mobility and/or the consideration of uncompensated charges in the transistor channel.
}


\usepackage{xspace}
\newcommand{\etal}{\textit{et~al.}\xspace}
\begin{document}


\section{Introduction}


The fabrication of organic thin-film transistors (TFTs) has reached a level at which devices with excellent performance, small device-to-device variations, and smooth electrical characteristics with low hysteresis are routinely available.\cite{guo_current_2017,paterson_recent_2018,yamamura_wafer-scale_2018,ogier_uniform_2018} These technological advances are significantly ahead of our current ability to reliably extract crucial transistor parameters, be that to design circuits, to determine material parameters, or to further optimize a device. The most prominent of these transistor parameters are the charge-carrier mobility as a material parameter and the contact resistance as an indicator for the quality of the metal-semiconductor interfaces. To be able to extract such parameters from the electrical device characteristics, the transistor operation and, hence, its electric characteristics must be understood in terms of these parameters.

In general, parameter extraction requires a theoretical model for the transistor operation that provides the current-voltage relations on the basis of input parameters that account for the point of operation (applied voltages), material properties and the device geometry. While material-related transistor parameters comprise, for example, the charge-carrier mobility and the gate-insulator permittivity, the most prominent geometry parameters are the length $L$ and the width $W$ of the transistor channel and the gate-insulator thickness. Such theoretical models hold the promise of being able to associate any changes in the current-voltage relation to changes in these parameters. Hence, it is particularly desirable to utilize a theoretical model that associates the drain current to these transistor parameters, preferably with a closed analytic expression. To obtain reliable and robust associations, it is customary to conceive specific models for each class of TFTs by accounting, for example, for a particular transport mechanism\cite{pasveer_unified_2005,li_electric-field_2017} or for particular geometry features, such as short channels.\cite{locci_modeling_2008} The potential success of a theoretical model inherently relies on preliminary assumptions that are guided by the device geometry and the anticipated transport mechanism. For instance, in the presumably most prominent model, the gradual channel approximation, it is assumed that all mobile charges are confined to the interface between the semiconductor and the gate insulator. Despite many efforts to improve the transistor models to better comply with the measured electrical characteristics,\cite{marinov_organic_2009,di_pietro_simultaneous_2014} the development of refined models is hampered as there is no reliable tool to check the consistency between the prediction made by a given theoretical model and the experimentally measured electrical characteristics.

Here we propose a technique to scrutinize the adequateness of the underlying theoretical model. The technique consists of a two-step process that requires a set of TFTs with different channel lengths. The two steps combine the benefits and overcome the drawbacks of the two classes of established extraction approaches, namely 'single transistor methods' and 'channel-length-scaling approaches'.\cite{natali_charge_2012} 'Single transistor methods' seek to extract the parameters of an assumed transistor model from certain voltage regions in the output or/and transfer characteristics of an individual TFT,\cite{natali_charge_2012,wang_transition-voltage_2010,takagaki_extraction_2016,torricelli_single-transistor_2014,di_pietro_simultaneous_2014} whereas in 'channel-length-scaling approaches' parameters are extracted from a series of nominally equivalent TFTs, that differ only in the channel length, by exploring the scaling of the transistor performance with the channel length from the perspective of the assumed model.\cite{kanicki_performance_1991,luan_experimental_1992,natali_modeling_2007} Neither of these two approaches is able to provide a reliable check of the consistency between theoretical model and measured current-voltage characteristics. For 'single transistor methods' the consistency can, at most, be checked within the limited region from which the parameters are extracted, and for 'channel-length-scaling approaches', the deviations of model predictions from the measured data is often hidden by device-to-device variations.

The technique we present here combines main aspects of the two classes of extraction methods. This combination allows us to go beyond extraction methods and enables a reliable check of the adequateness of the underlying theoretical model. Our first step analyzes single transistors. We fit the entire set of measured data points of all output and transfer characteristics at once to the assumed model. As pointed out by Deen \etal \cite{deen_organic_2009} and Fischer \etal,\cite{fischer_nonlinear_2017} the consideration of all available data points guarantees the best possible parameter set describing an individual transistor as a whole and eliminates the aforementioned ambiguity that arises from selecting certain regions of operation. The extracted parameter set is then used to calculate the corresponding output and transfer characteristics $I_{D}(V_{DS})$ and $I_{D}(V_{GS})$. Comparing the calculated electrical characteristics to the measured ones allows a first check of the validity of the assumed model.\cite{deen_organic_2009,fischer_nonlinear_2017} Furthermore, deviations seen in the characteristics can be analyzed to get an idea of how the model should be improved. If this check is successful and the characteristics match well, we can proceed to the second step and compare the results of the individually extracted parameters of all devices. The second step relies on the hypothesis that transistor quantities, such as voltage drops, resistances, and charge mobilities, can be split into contributions from the channel and from the contacts. If the assumed model correctly assigns the contributions to the channel and to the contacts, all channel-length dependencies are captured explicitly in the model. In turn, all related parameters have to be independent of the channel length. Hence, if, in a second check, the extracted parameters are found to be independent of the channel length, it can be concluded that the assumed model describes the measured devices consistently. The second step is of particular importance, because fitting approaches have the drawback that they can produce nice fits even for unreasonable models, provided that a sufficient number of parameters are considered.\cite{mayer_drawing_2010} As we overcome the drawbacks of both extraction methods and fitting approaches, our two-step fitting approach (TSFA) is suited for checking complex models and for identifying problems within those models.

We test the merit of our TSFA and scrutinize existing organic TFT models using experimental data. We purposefully select five sets of organic TFTs. These sets differ in the semiconductor, the geometry and the treatment of the semiconductor-contact interface to realize devices with nearly ideal (vanishing contact resistance) to highly non-ideal injection (large, non-linear contact resistances). In particular, we fabricate a set of bottom-gate, bottom-contact TFTs and bottom-gate, top-contact TFTs with dinaphtho[2,3-b:2’,3’-f]thieno[3,2-b]thiophene (DNTT) as the semiconductor and Au contacts. In the case of the bottom-gate, bottom-contact TFTs, the Au contact surfaces were functionalized with a layer of pentafluorobenzenethiol (PFBT) to reduce the contact resistance.\cite{gundlach_contact-induced_2008} The remaining transistors are bottom-gate, bottom-contact TFTs; one with pentacene as semiconductor and Au contacts functionalized with 2-phenylpyrimidine-5-thiol and two with C60 as the semiconductor and Au contacts functionalized with 4-(2-mercaptophenyl)pyrimidine and biphenyl-4-thiol respectively.\cite{petritz_embedded_2018} 
The DNTT-based bottom-gate, bottom-contact TFT set resembles an ideal transistor behaviour with low contact resistances very closely. 
Hence, this set will serve us as a reference and is analyzed in detail. 
First, we explain the application and interpretation of the most commonly used extraction method, the transmission line method (TLM),\cite{kanicki_performance_1991,luan_experimental_1992} in a step-by-step manner. Second, we illustrate our TSFA on the example of the model assumed in the TLM. Third, we test a more sophisticated model with field- and charge-carrier-density-dependent mobility. And finally, we test models with field- and charge-carrier-density-dependent mobility and non-linear contact resistances by analyzing the measured data of the other four TFT sets.

\section{Materials and Methods}\label{chap:materials_and_methods}

This chapter discusses in detail, (i), how to numerically calculate the drain current within the equivalent cirquit model we use, (ii), how we perform the fit of the calculated drain current to the measured data and, (iii), which transistor technologies we investigated with our TSFA.

\subsection{Equivalent Cirquit Model}

\begin{figure}[H]
\centering
\includegraphics[width=0.2\textwidth]{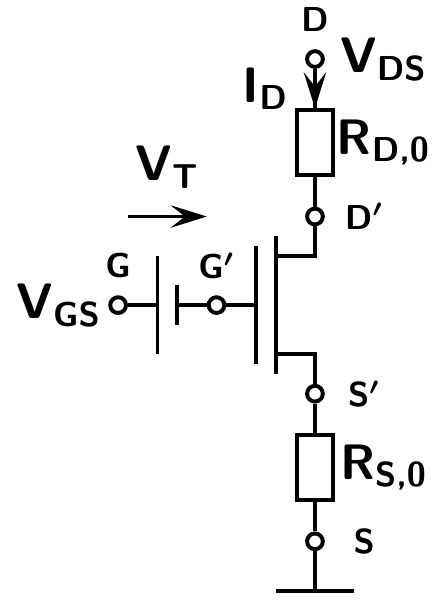}
\caption{Equivalent circuit model with an ideal transistor in the gradual channel approximation and a field- and charge-carrier-density-dependent mobility in the channel, connected to ohmic source and drain resistances $R_{S,0}$ and $R_{D,0}$. The threshold voltage $V_{T}$ is included as an external bias. The contacts of the ideal transistor are labeled source \textbf{S'}, drain \textbf{D'} and gate \textbf{G'} and the experimentally accessible contacts are labeled source \textbf{S}, drain \textbf{D} and gate \textbf{G}.}
\label{fig:model}
\end{figure}

The equivalent circuit model employed here is shown in Figure~\ref{fig:model}. This model contains an ideal transistor in the gradual channel approximation \cite{shockley_unipolar_1952} with a field- \cite{hall_poole-frenkel_1971,locci_modeling_2008} and charge-carrier-density-dependent mobility \cite{vissenberg_theory_1998,horowitz_temperature_2000} between the ideal source \textbf{S'}, drain \textbf{D'} and gate \textbf{G'} terminals. At the gate, the threshold voltage $V_{T}$ is considered as an external bias, and source and drain are connected to ohmic contact resistances $R_{S,0}$ and $R_{D,0}$. The experimentally accessible contacts are labeled source \textbf{S}, drain \textbf{D} and gate \textbf{G}.

The mobility $\mu$ at a certain position $x$ in the channel can be written as:
\begin{equation}
\mu(x)=\mu_{0}\exp\left(\beta\sqrt{\frac{L_{0}}{L}\left|\frac{V_{D'S'}}{V_{0}}\right|}\right)\left(\frac{V_{GS'}-V_{T}-V_{ChS'}(x)}{V_{0}}\right)^{\gamma}
\end{equation}
with the channel potential with respect to the source $V_{ChS'}(x)$ at this position $x$, the gate-source voltage $V_{G'S'}=V_{GS'}-V_{T}$, the drain-source voltage $V_{D'S'}$, the mobility prefactor $\mu_{0}$, the channel length $L$, the exponent of the field sensitivity $\beta$, the charge-carrier density sensitivity $\gamma$, and a constant length scale $L_{0}=1$~$\mu$m. To conveniently address both hole and electron conduction, we introduce a constant potential scaling factor $V_{0}$ with $V_{0}=1$~V for electron-conducting devices and $V_{0}=-1$~V for hole-conducting devices. Note that the absolute values of the constant length scale $L_{0}$ and the constant potential scale $V_{0}$ are chosen arbitrarily and are necessary only to avoid inconsistencies regarding the units within the corresponding power functions. The exponential term mimics a simplified Poole-Frenkel field-dependence\cite{hall_poole-frenkel_1971,locci_modeling_2008} and the right term describes the charge-carrier-density dependence with a power law behavior.\cite{vissenberg_theory_1998,horowitz_temperature_2000}

Incorporating the gradual channel approximation (for details see \cite{shockley_unipolar_1952,marinov_organic_2009}) leads to an implicit system of equations that determines the drain current $I_{D}$ for given applied gate-source and drain-source voltages $V_{GS}$ and $V_{DS}$:
\begin{eqnarray}
v_{G'S'}&=&\frac{1}{V_{0}}\left(V_{GS}-V_{T}-I_{D}\frac{r_{S,0}}{W}\right)\nonumber\\
v_{G'D'}&=&\frac{1}{V_{0}}\left(V_{GS}-V_{T}-V_{DS}+I_{D}\frac{r_{D,0}}{W}\right)\nonumber\\
I_{D}&=&\frac{V_{0}|V_{0}|W C_{I} \mu_{0}}{L(\gamma+2)}\exp\left(\beta\sqrt{\frac{L_{0}}{L}\left|v_{G'S'}-v_{G'D'}\right|}\right)\left[v_{G'S'}^{\gamma+2}\Theta(v_{G'S'})-v_{G'D'}^{\gamma+2}\Theta(v_{G'D'})\right]
\label{eq:implicit_drain_current}
\end{eqnarray}
The reduced voltages $v_{G'S'}$ and $v_{G'D'}$ are the voltages at the ideal gate \textbf{G'} to source \textbf{S'} and gate \textbf{G'} to drain \textbf{D'} contacts divided by $V_{0}$. The heaviside-function  $\Theta(x)$ is 1 for $x\ge 0$ and 0 for $x<0$. Furthermore, contact resistances $r = R W$ reduced by the channel width $W$, i.e., the source-sided $r_{S,0}=R_{S,0}W$ and the drain-sided $r_{D,0}=R_{D,0}W$ resistances, as well as the gate capacitance per unit area, $C_{I}$, are introduced. 

In summary, the drain current $I_{D}$ as output quantity is implicitly determined by two input quantities $V_{GS}$ and $V_{DS}$, six fit parameters $V_{T}$, $\mu_{0}$, $r_{S,0}$, $r_{D,0}$, $\beta$ and $\gamma$, two constants $L_{0}$ and $V_{0}$ and three geometry factors $L$, $W$ and $C_{I}$. The gate capacitance per unit area $C_{I}$ is a geometry factor, because it is approximately calculated from the thickness and the dielectric constant of the gate oxide.

The implicit system of equations \eqref{eq:implicit_drain_current} can be numerically solved with the bisection method incorporating knowledge of the desired fixed point. We start by setting $I_{D}^{(0)}=0$~A in the first two equations of \eqref{eq:implicit_drain_current} to get $v_{G'S'}^{(0)}$ and $v_{G'D'}^{(0)}$, and then substituting the latter in the right-hand side of the third equation. This gives $I_{D}^{(1)}$ and defines the search interval $[I_{D,min},I_{D,max}]=[\min(I_{D}^{(0)},I_{D}^{(1)}),\max(I_{D}^{(0)},I_{D}^{(1)})]$. Now the recurrent series starts by taking the midpoint $I_{D,MP}=(I_{D,min}+I_{D,max})/2$ and plugging it into the first two equations and the right side of the third equation of \eqref{eq:implicit_drain_current} to get $I_{D,calc}$. If $I_{D,MP}<I_{D,calc}$, the new search interval is $[I_{D,MP},\min(I_{D,max},I_{D,calc})]$ and if $I_{D,MP}>I_{D,calc}$, the new search interval is $[\max(I_{D,min},I_{D,calc}),I_{D,MP}]$. Calculating $I_{D,MP}$ and $I_{D,calc}$ is continued until the desired accuracy is reached.

\subsection{Fitting Procedure}

Fitting measured characteristics to this model is performed with a Gau\ss -Newton algorithm including the variation of Marquardt.\cite{marquardt_algorithm_1963} The algorithm has been modified slightly to be able to handle minimum and maximum values of parameters. In our case, $\mu_{0}$, $r_{S,0}$, $r_{D,0}$ and $\beta$ have to be positive and $\gamma >-1$.

The Gau\ss -Newton-Marquardt algorithm calculates the difference $\boldsymbol{\Delta a}=\boldsymbol{a}-\boldsymbol{a^{(0)}}$ between the current model parameters $\boldsymbol{a^{(0)}}$ and the suggested new model parameters $\boldsymbol{a}$ by solving the linear equation system
\begin{equation}
(A+\lambda D)\boldsymbol{\Delta a}=\boldsymbol{b}
\label{eq:marquardt_lin_eq_sys}
\end{equation}
with matrices $A$ and $D$, the convergence parameter $\lambda$ introduced by Marquardt and a vector $\boldsymbol{b}$. The matrix $A$ is given by
\begin{equation}
(A)_{ij}=\sum_{k=1}^{n}\frac{1}{\sigma_{k}^{2}}\frac{\partial I_{D}(V_{DS}^{(k)},V_{GS}^{(k)};\boldsymbol{a^{(0)}})}{\partial a_{i}}\frac{\partial I_{D}(V_{DS}^{(k)},V_{GS}^{(k)};\boldsymbol{a^{(0)}})}{\partial a_{j}}\text{,}
\end{equation}
containing the sum over all $n$ measured values $k$, the standard deviation $\sigma_{k}$ and the partial derivatives $\partial I_{D}(V_{DS}^{(k)},V_{GS}^{(k)};\boldsymbol{a^{(0)}})/\partial a_{i/j}$ of the calculated drain current $I_{D}$ at the measured data values $V_{DS}^{(k)}$ and $V_{GS}^{(k)}$ and the current model parameters $\boldsymbol{a^{(0)}}$ with respect to the model parameter $a_{i}$ and $a_{j}$, respectively. The matrix $D$ is a diagonal matrix consisting of the diagonal elements of $A$, $(D)_{ij}=\delta_{ij}(A)_{ij}$ with $\delta_{ij}$ being the Kronecker delta returning 1 if $i=j$ and 0 if $i\ne j$. The vector $\boldsymbol{b}$ is given by
\begin{equation}
b_i=\sum_{k=1}^{n} \frac{I_{D}^{(k)}-I_{D}(V_{DS}^{(k)},V_{GS}^{(k)};\boldsymbol{a^{(0)}})}{\sigma_{k}^{2}}\frac{\partial I_{D}(V_{DS}^{(k)},V_{GS}^{(k)};\boldsymbol{a^{(0)}})}{\partial a_{i}}
\end{equation}
involving the measured drain current $I_{D}^{(k)}$ corresponding to the measured voltages $V_{DS}^{(k)}$ and $V_{GS}^{(k)}$.

To consider minimum and maximum values of model parameters, the matrices $A$ and $D$, the vector $b$ and the convergence parameter $\lambda$ are evaluated as in Ref.~\cite{marquardt_algorithm_1963} and the linear equation system \eqref{eq:marquardt_lin_eq_sys} is solved to receive $\boldsymbol{\Delta a}$. Before going on with this calculated value for $\boldsymbol{\Delta a}$, it is checked if any of the suggested parameters $\boldsymbol{a}=\boldsymbol{a^{(0)}}+\boldsymbol{\Delta a}$ are out of bounds. If this is the case, the corresponding value for $\Delta a_{j}$ of the entry $j$ that is allowed to stay within the boundaries is calculated (e.g., $\Delta a_{j}=a_{j}^{max}-a_{j}^{(0)}$ if the upper boundary is exceeded) and plugged into the linear equation system \eqref{eq:marquardt_lin_eq_sys} by eliminating the corresponding equation $j$ and transferring $(A)_{ij}\Delta a_{j}$ to the right side $b_i\rightarrow b_i-(A)_{ij}\Delta a_{j}$. The new linear equation system is solved and the model parameters are checked again. This procedure is iteratively continued until all model parameters are in bounds. Following this, the Gau\ss -Newton algorithm is continued.

To calculate the required derivatives of the model function with respect to the model parameters, a few definitions are useful:
\begin{eqnarray}
T_0&=&\beta\sqrt{\frac{L_0}{L}}\frac{v_{G'S'}^{\gamma+2}\Theta(v_{G'S'})-v_{G'D'}^{\gamma+2}\Theta(v_{G'D'})}{2(\gamma+2)\sqrt{\left|v_{G'S'}-v_{G'D'}\right|}}\text{sgn}\left(v_{G'S'}-v_{G'D'}\right) \text{,}\\
T_{G'S'}&=&v_{G'S'}^{\gamma+1}\Theta(v_{G'S'})+T_0\text{,}\\
T_{G'D'}&=&v_{G'D'}^{\gamma+1}\Theta(v_{G'D'})+T_0\text{,}\\
\tilde{\mu}_0&=&\mu_{0}\exp\left(\beta\sqrt{\frac{L_{0}}{L}\left|v_{G'S'}-v_{G'D'}\right|}\right)\text{,}\\
D_{I_{D}}&=&1+\frac{\left|V_{0}\right|C_{I}\tilde{\mu}_0}{L}\left(T_{G'S'}r_{S,0}+T_{G'D'}r_{D,0}\right)
\end{eqnarray}
The sign function $\text{sgn}(x)$ is -1 if $x<0$, 1 if $x>0$ and 0 if $x=0$. With these definitions, the derivatives can be written in a compact way:
\begin{eqnarray}
\frac{\partial I_{D}}{\partial V_{T}}&=&-\frac{|V_{0}|WC_{I}\tilde{\mu}_0}{LD_{I_{D}}}\left(T_{G'S'}-T_{G'D'}\right)\\
\frac{\partial I_{D}}{\partial \mu_{0}}&=&\frac{I_{D}}{\mu_{0}D_{I_{D}}}\\
\frac{\partial I_{D}}{\partial r_{S,0}}&=&-\frac{|V_{0}|C_{I}\tilde{\mu}_{0} T_{G'S'}I_{D}}{LD_{I_{D}}}\\
\frac{\partial I_{D}}{\partial r_{D,0}}&=&-\frac{|V_{0}|C_{I}\tilde{\mu}_{0} T_{G'D'}I_{D}}{LD_{I_{D}}}\\
\frac{\partial I_{D}}{\partial \gamma}&=&-\frac{I_{D}}{D_{I_{D}}(\gamma+2)}\nonumber\\
&&-\frac{V_{0}\left|V_{0}\right|WC_{I}\tilde{\mu}_{0}}{L(\gamma+2)D_{I_{D}}}\left[\ln(v_{G'S'})v_{G'S'}^{\gamma+2}\Theta(v_{G'S'})-\ln(v_{G'D'})v_{G'D'}^{\gamma+2}\Theta(v_{G'D'})\right]\\
\frac{\partial I_{D}}{\partial \beta}&=&\frac{I_{D}}{D_{I_{D}}}\sqrt{\frac{L_0}{L}\left|v_{G'S'}-v_{G'D'}\right|}
\end{eqnarray}
In addition to these derivatives, starting values for the fitting procedure are required. Initially, we can set all parameters to zero except the mobility prefactor $\mu_{0}$ and the threshold voltage $V_{T}$. These two parameters can be estimated from the saturation regime of the output characteristics. In this regime with only $\mu_{0}$ and $V_{T}$ being non-zero, the drain current $I_{D}$ is calculated by $I_{D,sat}=WC_{I}\mu_{0}(V_{GS}-V_{T})^2/2L$. Performing a linear fit of $\sqrt{I_{D,sat}(V_{GS})}$ provides starting values for $\mu_{0}$ and $V_{T}$. With these starting values, the first fit is performed by optimizing only $\mu_0$ and $V_{T}$. Starting from these optimized parameters, more and more parameters are included in the fitting procedure. The next fit, e.g., is optimizing $\mu_{0}$, $V_{T}$, $r_{S,0}$ and $r_{D,0}$ followed by a fit of $\mu_{0}$, $V_{T}$, $r_{S,0}$, $r_{D,0}$ and $\gamma$ and a final fit of $\mu_{0}$, $V_{T}$, $r_{S,0}$, $r_{D,0}$, $\gamma$ and $\beta$. When changing the order of included fit parameters (e.g. $\beta$ before $\gamma$), the optimized parameters should converge to the same solution within the chosen numerical accuracy.

\subsection{Fabricated Devices}

All TFTs were fabricated on flexible plastic substrates and share aluminum oxide as gate dielectric layer.
The TFTs investigated in particular detail are bottom-gate, bottom-contact TFTs with 
a 30~nm thick layer of 
DNTT as the semiconductor and Au contacts that are treated with PFBT to increase the work function of the contacts\cite{hong_tuning_2008} and to improve the semiconductor morphology across the contact interface\cite{gundlach_contact-induced_2008}.
The ultrathin 5.3 nm aluminum oxide gate dielectric layer enables operation voltages below 3~V.\cite{borchert_small_2019}
This set of TFTs was chosen because it appears to closely resemble an ideal transistor, as demonstrated by a nearly perfect linear behavior in the linear regime of the output characteristics, low contact resistances, and good reproducibility. This nearly ideal behavior is maintained even for the smallest channel length of $L=2$~$\mu$m.

The remaining sets of TFTs, that were analyzed for comparison, are a series of bottom-gate, top-contact TFTs\cite{ante_contact_2012} and series of bottom-gate, bottom-contact TFT with either pentacene or C60 as the semiconductor and Au contacts decorated with biphenyl-based SAMs containing embedded dipoles (one phenyl ring exchanged by pyrimidine) to adjust the work function of the contacts.\cite{petritz_embedded_2018}




\section{Results}


\subsection{Transmission Line Method}

Before our TSFA is applied, we analyze the data measured for our set of DNTT-based bottom-gate, bottom-contact TFTs with the widely used transmission line method (TLM). This analysis is performed to (i) put the measured data into a perspective commonly shared in our field of research and (ii) highlight the benefits and drawbacks of the TLM.

In principle, the TLM is able to take into account non-idealities like non-ohmic contact resistances. However, when applying the most common TLM extraction procedure, the model assumptions are rather strict, as it assumes ideal transistors that satisfy the gradual channel approximation \cite{shockley_unipolar_1952} and have a constant mobility and ohmic source and drain resistances.\cite{kanicki_performance_1991,luan_experimental_1992} With these model assumptions, the drain current $I_{D}$ in the linear regime of the output characteristics can be written as
\begin{equation}
I_{D}=\frac{V_{0}WC_{I}\mu}{2\left|V_{0}\right|\left(L+L_{T}\right)}\left[\left(V_{GS}-V_{T}-I_{D}\frac{r_{S,0}}{W}\right)^{2}-\left(V_{GS}-V_{T}-V_{DS}+I_{D}\frac{r_{D,0}}{W}\right)^{2}\right]\text{.}
\end{equation} 
The transfer length $L_{T}$ accounts for a channel-length-independent extension of the channel in the contact regions. In bottom-gate, top-contact TFTs, $L_{T}$ can be interpreted as the additional distance that charge carriers have to travel through the semiconductor to reach the channel (see e.g. \cite{fischer_nonlinear_2017}). For bottom-gate, bottom-contact TFTs, charges are injected very close to the channel and travel significantly a shorter distance through the semiconductor before reaching the channel. This implies that $L_{T}$ by its own is not a physically interpretable parameter but rather has to be seen as a weighting factor for a non-ohmic contribution to the contact resistance.

\begin{figure}[H]
\centering
\includegraphics[width=0.75\textwidth]{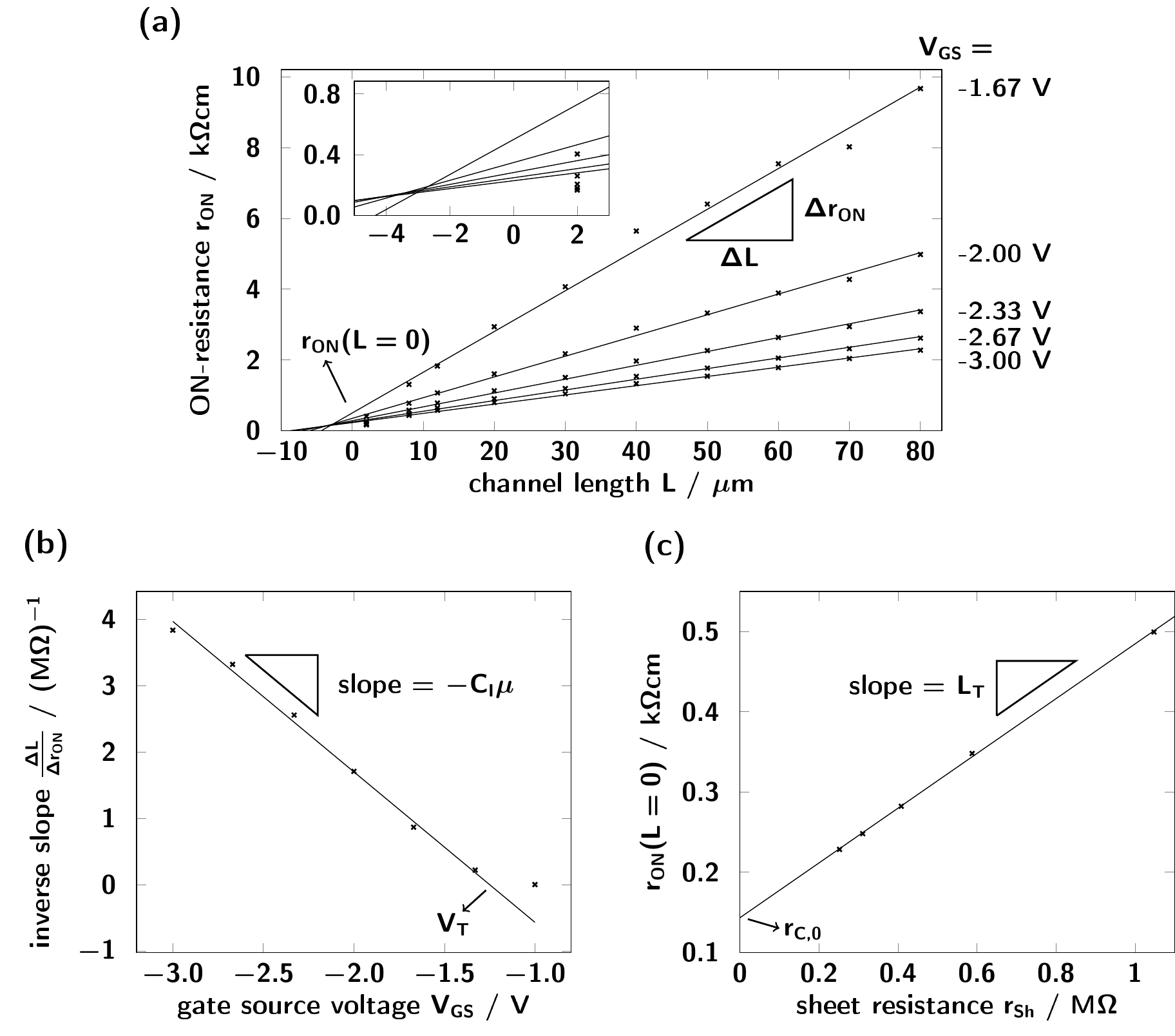}
\caption{Parameter extraction in the framework of the transmission line method (TLM), performed here on bottom-gate, bottom-contact TFTs based on the small-molecule semiconductor DNTT. In (\textbf{a}), the ON-state resistance $r_{ON}=W\partial V_{DS}/\partial I_{D}$ for $V_{DS}\rightarrow 0$~V, extracted from the measured output characteristics, is plotted as a function of the channel length for different gate-source voltages $V_{GS}$. From a linear fit of these data points, the inverse slope $\Delta L/\Delta r_{ON}$ and the y-axis intercept $r_{ON}(L=0)$ are extracted. The insert shows a magnification of the intercept of all fit lines and the extracted $r_{ON}$ values for the smallest channel length (symbols). In (\textbf{b}), $\Delta L/\Delta r_{ON}$ plotted versus $V_{GS}$ yields the threshold voltage $V_{T}=1.25$~V and the intrinsic channel mobility $\mu=3.2$~cm$^{2}$/Vs. In (\textbf{c}), $r_{ON}(L=0)=r_{Sh}L_{T}+r_{C,0}$ is plotted versus the sheet resistance $r_{Sh}=\left|V_{0}\right|[V_{0}C_{I}\mu(V_{GS}-V_{T})]^{-1}$ to obtain the transfer length $L_{T}=3.4$~$\mu$m and the total ohmic contact resistance $r_{C,0}=0.14$~$\Omega$cm.}
\label{fig:TLM}
\end{figure}

The parameter extraction procedure consists of three parts. 
In the first part, the ON-state resistance $r_{ON}$ is calculated from the slope of the measured output characteristics:
\begin{equation}
r_{ON}=\lim_{V_{DS}\rightarrow 0} W\frac{\partial V_{DS}}{\partial I_{D}}=\left|V_{0}\right|\frac{L+L_{T}}{V_{0}C_{I}\mu\left(V_{GS}-V_{T}\right)}+r_{C,0}
\end{equation}
with $r_{C,0}=r_{S,0}+r_{D,0}$. Note that it is important to extract $r_{ON}$ for $V_{DS}\rightarrow 0$~V because only at this point a clear separation of contact and channel is possible within the model (cf. supplementary materials Figure \textbf{S1}). To determine $r_{ON}$, we performed a linear fit for the four smallest measured drain-source voltages and forced this fit to go through the origin $V_{DS}=0$~V and $I_{D}=0$~A. The plot of $r_{ON}$ as a function of the channel length $L$ for different $V_{GS}$ is shown in Figure~\ref{fig:TLM}(\textbf{a}). The measured $r_{ON}$ behaves linearly with respect to $L$ and the intercept of all curves for different gate-source voltages at the bottom left is approximately at $L\approx -3.2$~$\mu$m and $r_{ON}\approx 0.15$~k$\Omega$cm.

In the second part, the inverse slope $\Delta L/\Delta r_{ON}=C_{I}\mu\left(V_{GS}-V_{T}\right)V_{0}/\left|V_{0}\right|$ is extracted from Figure~\ref{fig:TLM}(\textbf{a}) and plotted versus $V_{GS}$ (see Figure~\ref{fig:TLM}(\textbf{b})). The slope of this graph yields the intrinsic channel mobility $\mu=3.2$~cm$^{2}$/Vs and the x-axis intercept gives the threshold voltage $V_{T}=-1.25$~V. In the last part of parameter extraction procedure, the ON-state resistance at zero channel length $r_{ON}(L=0)=r_{Sh}L_{T}+r_{C,0}$ is plotted as a function of the sheet resistance $r_{Sh}=\left|V_{0}\right|[V_{0}C_{I}\mu(V_{GS}-V_{T})]^{-1}$ (see Figure~\ref{fig:TLM}(\textbf{c})). The slope from the linear fit of this data is the transfer length $L_{T}=3.4$~$\mu$m and the y-axis intercept yields the ohmic contact resistance of $r_{C,0}=0.14$~k$\Omega$cm.

\begin{figure}[H]
\centering
\includegraphics[width=0.95\textwidth]{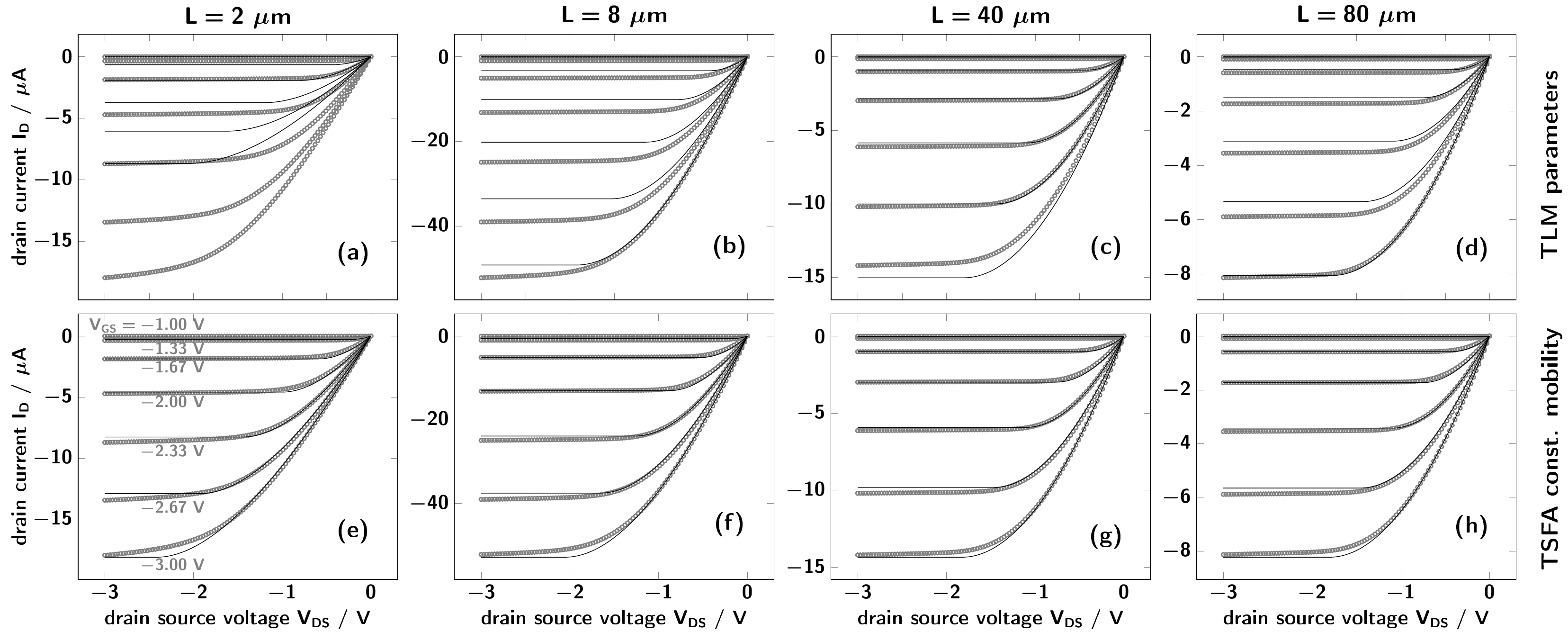}
\caption{Measured (gray symbols) and calculated (black lines) output characteristics for different channel lengths of DNTT-based bottom-gate, bottom-contact TFTs (corresponding transfer characteristics, see supplementary materials Figure \textbf{S2}). 
Note that the symbols appear as an apparent thick line due to the close spacing of the voltage points.
The calculated curves in the first row were obtained using the parameters extracted using the TLM, whereas the results in the second row were calculated using the TSFA for the model used within the TLM.}
\label{fig:TLM_FA1_OC}
\end{figure}

To check the reliability of the parameters extracted by the TLM, the following requirements must be fulfilled:
\begin{itemize}[leftmargin=*,labelsep=5.8mm]
\item Looking at $V_{DS}\rightarrow 0$~V of the measured output characteristics (cf. gray symbols in Figure~\ref{fig:TLM_FA1_OC}), the curves must show a linear onset and the slope must monotonically decrease with increasing $V_{DS}$. An S-shape of the curves in this region is a clear indicator for a non-ohmic contact resistance.
\item The measured data must be represented by the linear fits for all three cases $r_{ON}$ versus $L$, $\Delta L/\Delta r_{ON}$ versus $V_{GS}$ and $r_{ON}(L=0)$ versus $r_{Sh}$.
\item The transfer length $L_{T}$ and the total contact resistance $r_{C,0}$ must be equal to the intercept of the $r_{ON}(L)$ curves for different $V_{GS}$.
\end{itemize}

For the set of TFTs analyzed in Figure~\ref{fig:TLM}, all of the above requirements are indeed met. Small deviations of the extracted $r_{ON}$ values for different channel lengths from the linear fit (see Figure~\ref{fig:TLM}(\textbf{a})) can be attributed to device-to-device variations. 
A closer look, however, reveals inconsistencies.
The inset in Figure~\ref{fig:TLM}(\textbf{a}) shows a magnification of $r_{ON}$ versus $L$ close to $L=0$ together with the $r_{ON}$ values for the smallest channel length $L=2$~$\mu$m (crosses).
As can be seen, the fit lines do not cross all in one point. In addition,  $r_{ON}$ of the TFT with the smallest channel length $L=2$~$\mu$m is always a factor of approximately two below the linear fit. 
Both inconsistencies do not prevent a further analysis, because the deviation of the $L=2$~$\mu$m TFT might be due to short-channel effects, while the fact that the fit lines do not cross in one point could be a consequence of the drain-source voltage being too large to extract the ON-resistance in a reliable manner (cf. supplementary materials Figure \textbf{S1}). 
These explanations do not necessarily affect the validity of the model system.

As we are able to calculate characteristics for given parameters, we can compare output characteristics calculated with the parameters extracted using the TLM to the measured output characteristics.
This comparison is shown in the first row of Figure~\ref{fig:TLM_FA1_OC} for different channel lengths. 
As can be seen, the calculated curves (black lines) deviate substantially from the measured curves (gray symbols), regardless of the channel length.
These deviations indicate a problem within the TLM that was not spotted by the reliability check performed above. 
Upon closer inspection, it can be noticed that the calculations match the experimental data better for longer channel lengths. The curves for the devices with the largest channel length $L=80$~$\mu$m (see Figure~\ref{fig:TLM_FA1_OC}(\textbf{d})) and also for the intermediate channel lengths $L=40$~$\mu$m (see Figure~\ref{fig:TLM_FA1_OC}(\textbf{c})) and $L=8$~$\mu$m (see Figure~\ref{fig:TLM_FA1_OC}(\textbf{b})) show at least a reasonably good match, whereas in Figure~\ref{fig:TLM_FA1_OC}(\textbf{a}) the drain current is by far too small for the device with the smallest channel length $L=2$~$\mu$m. For the three longer channel lengths, the slope at the beginning of the linear regime is captured quite well, while the match becomes increasingly worse upon increase of $V_{DS}$ into the saturation regime. The better agreement in the linear regime is related to the fact that the parameters in the TLM are extracted from the slope at  $V_{DS}\rightarrow 0$~V.

\subsection{TSFA with Constant-Mobility Model}

One weakness of using the TLM to extract parameters is that all parameters have to be the same for all TFTs within the set of different channel lengths.
However, those parameters can vary considerably even for nominally equivalent TFTs. 
Then, device-to-device variations would potentially be able to explain the deviations of the measured and calculated output characteristics. So the question arises, whether the deviations can be attributed to the extraction method (TLM) or to the underlying transistor model.
To answer this question, we analyzed the measured TFT data with our TSFA. We extract an effective mobility  $\mu_{eff}$, threshold voltage $V_{T}$, source resistance $r_{S,0}$ and drain resistance $r_{D,0}$ for each TFT individually. 
The calculated output characteristics of these fits can be seen in the second row in Figure~\ref{fig:TLM_FA1_OC}. 
For all channel lengths, the calculated curves have notably improved compared to the ones referring to the TLM. As only minor deviations can be spotted, the important information taken from those curves is that the first step of our TSFA is conditionally passed. The details of the deviations between the measured and calculated output characteristics are discussed after the completion of the second step below. 

\begin{figure}[H]
\centering
\includegraphics[width=0.95\textwidth]{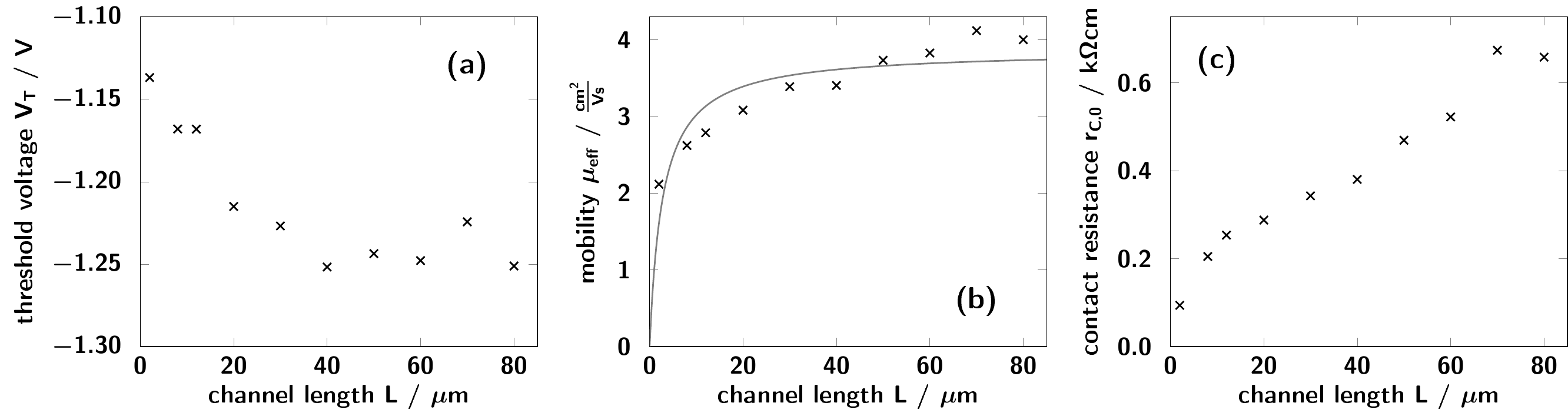}
\caption{Channel-length dependence of the parameters extracted with the TSFA for the model used within the TLM. The variation of the threshold voltage $V_{T}$ in (\textbf{a}) shows only minor $L$ dependence. For the mobility in (\textbf{b}), the appearing $L$ dependence (crosses) can not be consistently described by a transfer length $L_{T}$ with the corresponding fit $\mu_{eff}=\mu_{intr}\frac{L}{L+L_{T}}$ (solid line) and for the contact resistance $r_{C,0}=r_{S,0}+r_{D,0}$ in (\textbf{c}), the distinct linear increase with $L$ can not be explained at all. As a consequence, the model does not pass the second step.}
\label{fig:FA1_parameters}
\end{figure}

For the second step, we have to plot the extracted parameters versus the channel length, as shown in Figure~\ref{fig:FA1_parameters}. 
To be consistent with the model assumptions, these parameters need to be independent of $L$. 
In Figure~\ref{fig:FA1_parameters}(\textbf{a}), the threshold voltage $V_{T}$ exhibits a minor dependence on the channel length $L$ with an increase of about 100~mV for the smallest channel length. 
In Figure~\ref{fig:FA1_parameters}(\textbf{b}), a clear $L$ dependence of the effective mobility $\mu_{eff}$ (symbols) can be seen. 
If we strictly stick to the model underlying the TLM, we could surmise that this dependence could be related to the transfer length $L_{T}$. 
To check whether the introduction of a transfer length conceptually lifts the $L$ dependence,
we can incorporate $L_T$ into the second step by replacing $\mu_{eff}$ by $\mu_{intr}\frac{L}{L+L_{T}}$.
Then, the value of the intrinsic mobility $\mu_{intr}$ should be constant.\cite{rodel_contact_2013}
A fit of $\mu_{eff}=\mu_{intr}\frac{L}{L+L_{T}}$ is shown as a solid line in Figure~\ref{fig:FA1_parameters}(\textbf{b}). The shape of this fit does not represent the extracted parameter $\mu_{eff}$ well because it systematically overestimates the extracted parameters for intermediate channel lengths and underestimates them for high channel lengths. 
This poor match of the shapes indicates a problem with the model system. 
The right panel Figure~\ref{fig:FA1_parameters}(\textbf{c}) displays the combined contact resistance $r_{C,0}=r_{S,0}+r_{D,0}$. Rather than being independent of the channel length, the contact resistance $r_{C,0}$ grows by more than a factor of three with increasing $L$. This is a clear indicator for an inadequate transistor model.

To find the reason for the failure of the model, a closer look at the deviations of the calculated output characteristics from the measured ones can give an idea (see second row in Figure~\ref{fig:TLM_FA1_OC}). 
The deviations occur as two distinct symptoms.
First, the shape of the calculated curves at the transition between linear and saturation regime does not really fit to the measured data and second, the measured data shows a linear trend in the saturation regime which is not captured by the calculated curves. 
The first symptom appears regardless of the channel length and can be diminished by assuming a charge-carrier-density-dependent mobility of the form $\mu=\mu_{0}(V_{G}-V_{Ch})^{\gamma}$ as suggested by percolation theory \cite{vissenberg_theory_1998} or multiple trapping and release \cite{horowitz_temperature_2000}. 
The second symptom is more pronounced for shorter channels indicating a field-dependence of the mobility. As a first attempt, we assume a simplified Poole-Frenkel behavior of the form $\exp(\beta\sqrt{V_{DS}/L})$.\cite{hall_poole-frenkel_1971,locci_modeling_2008}

\subsection{TSFA with Field- and Charge-Carrier-Density-Dependent Mobility}

\begin{figure}[H]
\centering
\includegraphics[width=0.95\textwidth]{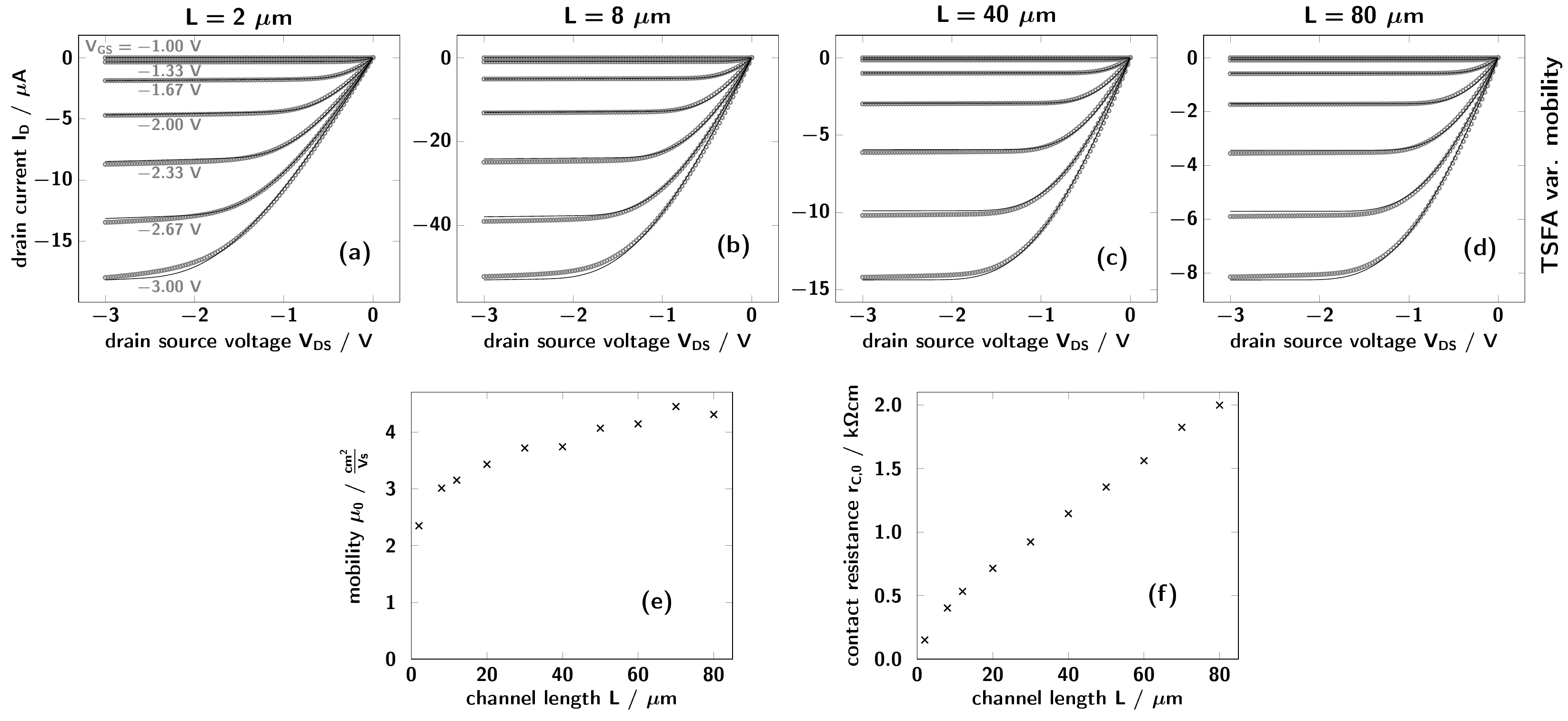}
\caption{Results of the TSFA for the model with field- and charge-carrier-density-dependent mobility. In (\textbf{a}) to (\textbf{d}), output characteristics for different channel lengths indicate a good agreement of the measured data (gray symbols) and the fit (black lines). Corresponding transfer characteristics are found in the supplementary materials Figure \textbf{S2}. In (\textbf{e}) and (\textbf{f}), the channel-length dependence of the mobility prefactor $\mu_{0}$ and the contact resistance $r_{C,0}=r_{S,0}+r_{D,0}$ indicate a failure of the model to properly represent the TFTs.}
\label{fig:FA2_OC_parameters}
\end{figure}

Incorporating a field- and charge-carrier-density-dependent mobility in the model leads to a clear improvement of the deviations between the measured and calculated output characteristics (see Figure~\ref{fig:FA2_OC_parameters}(\textbf{a}) to (\textbf{d})). Especially the TFT with the smallest channel length shows a much better agreement due to the improved description of the saturation regime with the Poole-Frenkel behavior. For all channel lengths, the curves of the more positive gate-source voltages $V_{GS}>-2.5$~V fit nearly perfectly. We again move on to examine the $L$ dependence of the extracted parameters. The most relevant parameters are the mobility prefactor $\mu_{0}$ and the combined contact resistance $r_{C,0}=r_{S,0}+r_{D,0}$ shown in Figure~\ref{fig:FA2_OC_parameters}(\textbf{e}) and (\textbf{f}). The mobility prefactor $\mu_{0}$ exhibits a slightly lower $L$ dependence compared to the effective mobility $\mu_{eff}$ examined earlier (cf. Figure~\ref{fig:FA1_parameters}(\textbf{b})). The $L$ dependence of $r_{C,0}$ is even more pronounced with approximately one order of magnitude between smallest and largest channel length (see Figure~\ref{fig:FA2_OC_parameters}(\textbf{f})), provoking a failure of this model. 
To illustrate the significant influence of the length-dependence of the contact resistance, Figure~\textbf{S3} (in the supplementary material) shows the disagreement of measured and calculated output characteristics when taking the contact resistance of the device with the smallest channel length (shown in Figure~\textbf{S3} (\textbf{a}) to (\textbf{d})) and the largest channel length (shown in Figure~\textbf{S3} (\textbf{e}) to (\textbf{h})). The remaining parameters, $V_{T}$, $\gamma$ and $\beta$ do not have such a pronounced $L$ dependence (not shown).

To identify the problem of the model, we can have a look at the output characteristics of all channel lengths, Figure~\ref{fig:FA2_OC_parameters}(\textbf{a}) to Figure~\ref{fig:FA2_OC_parameters}(\textbf{d}). In the saturation regime, the calculated curve for $V_{GS}=-2.67$~V always lies above the measured data and the calculated curve for $V_{GS}=-3.00$~V always lies below the measured data. This wrong spacing of the curves in the saturation regime is an indicator for a problem of the charge-carrier-density dependence of the mobility, which is predominantly determined by the gate-source voltage, $V_{GS}$.

The spacing of the curves in the saturation regime is not only determined by the charge-carrier-density dependence of the mobility, but also by the contact resistances (explained in more detail in the supplementary material, Figure~\textbf{S4}). Assuming a constant mobility and no contact resistance, the saturation current $I_{D,sat}$ increases quadratically with the gate-source voltage, $(V_{GS}-V_{T})^2$. On the other hand, assuming a constant mobility and a very high contact resistance, the saturation current would increase linearly with the gate-source voltage. This means that increasing both, mobility and contact resistance, can lead to similar $I_{D}(V_{DS})$ curves for the highest $V_{GS}$ and different spacing for lower $V_{GS}$ (see Figure~\textbf{S4}).

This effect could possibly explain the increase of $r_{C,0}$ with $L$ in the following way. If the charge-carrier-density dependence of the mobility is captured incorrectly, the spacing of the output characteristics for different gate-source voltages $V_{GS}$ will be wrong as well. The spacing is corrected by way of a compensating, though incorrect, change in the contact resistance. As the error of the mobility scales with $L$ in the calculation of the drain current because it is a channel property, and the contact resistance has no $L$ scaling effect, the extracted value of the contact resistance is forced to scale with $L$ to compensate the mobility.

The over- and underestimation of $I_{D}$ for the second lowest and lowest $V_{GS}$, respectively, suggests that the contact resistance tries to reduce the spacing for higher $V_{GS}$ and, hence, is too high. This change in spacing could be achieved as well if the mobility would decrease with increasing charge-carrier density. This decrease should only happen for high charge-carrier densities, because for low charge-carrier densities, related to low $V_{GS}$, the increasing mobility of the improved TFT model describes the measured curves much better than the constant mobility model. So the evaluation of our TSFA suggests that the mobility should first increase and later decrease with increasing charge-carrier density. Experimental hints indicating such a behavior of the mobility were recently found by Bittle \etal \cite{bittle_mobility_2016} and Uemura \etal \cite{uemura_extraction_2016}; Fishchuk \etal \cite{fishchuk_analytic_2007} suggested such a behavior from a theoretical point of view.

Besides improving the mobility, a potential alternative problem in the transistor model is that the gradual channel approximation disregards the fact that organic semiconductors are in principle insulators. As a consequence, all mobile charge carriers have to be brought externally into the channel. This charge accumulation is not compensated by charges of opposite polarity, in contrast to conventional semiconductors. This uncompensated charge accumulation affects the electric field at the contact with increasing impact for increasing channel length. Including this charge cloud in the transistor model might also be able to diminish the $L$ dependence of the contact resistance.

\subsection{Testing Additional TFT Technologies}

We note that the failure of the transistor model illustrated above is not a peculiarity of the chosen experimental TFT technology. Neither changing the geometry, nor the organic semiconductor, helps to improve the applicability of this transistor model. To confirm this claim, four more transistor technologies are investigated. These other technologies include a similar TFT set as above, only the Au contacts were changed from bottom-contact to top-contact while the thickness of the DNTT layer was kept at 30~nm (called DNTT - TC in the following).\cite{zschieschang_dinaphtho[23-b:23-f]thieno[32-b]thiophene_2011} In addition, three other bottom-gate, bottom-contact TFT series were examined: pentacene on Au contacts coated with a SAM of 2-phenylpyrimidine-5-thiol (Pentacene - BP0-down), C60 with 4-(2-mercaptophenyl)pyrimidine (C60 - BP0-up) and C60 with biphenyl-4-thiol (C60 - BP0) (for detail, see \cite{petritz_embedded_2018}). For DNTT - TC and C60 - BP0, non-linearities in the linear regime of the output characteristics were modeled with a gate-voltage-dependent Schottky diode at the source side to get a reasonable agreement of measured and fitted characteristics (for details about the Schottky diode, see \cite{petritz_embedded_2018}).

\begin{figure}[H]
\centering
\includegraphics[width=0.95\textwidth]{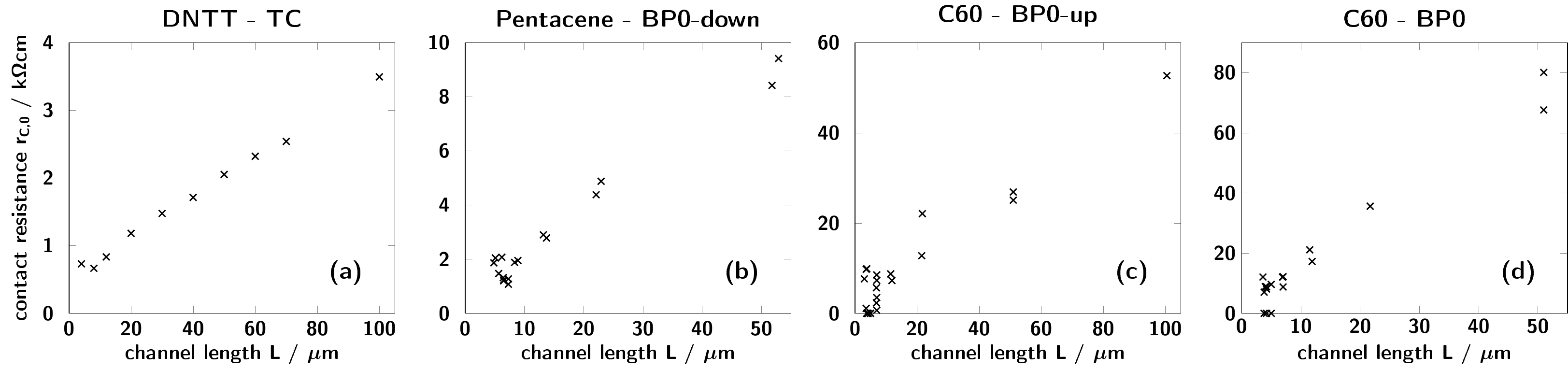}
\caption{Ohmic part of the contact resistance $r_{C,0}$ versus the channel length for different device series, i.e., a DNTT-based bottom-gate, top-contact TFT (\textbf{a}) and bottom-gate, bottom-contact TFTs with Pentacene on Au/BP0-down, C60 on Au/BP0-up and C60 on Au/BP0 contacts, respectively (\textbf{b}) to (\textbf{d}).
Despite different geometries, organic semiconductors, and contact preparations, all series exhibit a clear channel-length dependence of the ohmic part of the contact resistance $r_{C,0}$ . 
This leads to a failure of the transistor model in all instances.
The substantial fluctuations of $r_{C,0}$ for low $L$-values for the two C60 series (including transistors with $r_{C,0}=0$~k$\Omega$cm) reflects the fact that the uncertainties of the ohmic contact resistance for those channel lengths is in the order of the actual value. This high uncertainty does not obscure the clear increase of $r_{C,0}$ with $L$.}
\label{fig:FA_TC_PEN_C60BP0up_C60BP0}
\end{figure}

Figure~\ref{fig:FA_TC_PEN_C60BP0up_C60BP0} shows the ohmic part of the contact resistance $r_{C,0}$ as a function of the channel length $L$ for all of the additional four TFT series. The approximately linear dependence of $r_{C,0}$ on $L$ results in a similar failure of the transistor model in the second step of our TSFA for each device series.

\section{Summary and Conclusions}

In this paper, we propose a two-step fitting approach (TSFA) to check whether a transistor model is capable of describing the experimental characteristic of TFT devices. 
Only a valid transistor model, that correctly discriminates between contact and channel properties, enables one to reliably extract, interpret, and compare contact resistances and channel mobilities of TFTs.
The TSFA relies on a series of transistors with varying channel length and consists of two steps.
First, the chosen transistor model is fitted to all measured data points of output and transfer characteristics of each TFT separately to extract the transistor parameters of each device. Second, one checks whether the extracted parameters depend on the channel length. 
The latter consistency check is successful if (i) the measured data is represented well by the current-voltage curves calculated with the model and the transistor parameters and (ii) the extracted parameters are independent of the channel length. 
Our approach offers a clear benefit compared to currently used extraction methods, i.e., the reliability of the tested model can be easily checked. Due to the investigation of each individual TFT as a whole, the reason for a failure of the transistor model can be identified from the nature of the deviations between the measured data and the curves calculated with the extracted parameters.

We line out the indicators that are available to judge consistency within the TSFA by using the transistor model underlying the transmission line method (TLM) as an illustrative example.
TFTs with particularly small contact resistances served as test set, i.e., TFTs whose operation resemble the ideal transistor behavior as closely as possible. 
This test set readily exemplifies, that inconsistencies cannot be necessarily spotted within the parameter extraction step, but rather require a second step for validity checking.
An analysis with the TLM of the test set gave, at the first glance, an apparently consistent picture comprising (i) a linear onset of the output characteristics for zero drain-source voltage and (ii) a high quality of all performed linear fits to the corresponding data points. 
However, the transistor characteristics calculated with the extracted parameters failed to reproduce the measured curves. The subsequent validity check of the TSFA for the model assumed in the TLM was not passed, because the extracted contact resistances retained a pronounced dependence on the channel length. 
Such inconsistencies ought to be removed or, at least diminished, by improved transistors models.  For example, the model underlying the TLM can be improved by accounting for a field- and charge-carrier-density-dependent mobility.\cite{hall_poole-frenkel_1971,locci_modeling_2008,vissenberg_theory_1998,horowitz_temperature_2000}
Even though the TSFA attests better agreement between measured and calculated characteristics, also this improved model fails the subsequent validity check of the TSFA due to a marked remnant channel-length dependence of the contact resistance. 
The failure of the advanced transistor model featuring a field- and charge-carrier-density-dependent mobility was demonstrated for a broad selection of transistors, i.e., TFTs in a top-contact architecture, with different organic semiconductors, and high injection barriers that resulted in profound non-linear contributions to the contact resistance.

To improve the currently available transistor models, we need to face two aspects. 
On the one hand, the analysis of the deviations of the measured and calculated characteristics suggests that the charge-carrier-density dependence of the mobility is not captured correctly. Hence, a mobility model that is particularly suitable for the predominantly two-dimensional charge transport through the channel of a thin-film transistor has to be developed. On the other hand, the gradual channel approximation should be reconsidered by accounting for the charge accumulation in the channel, whose effect on the electric field distribution is not compensated by charges of opposite polarity within the organic semiconductor.
Our TSFA can be used to check each stage of model improvement.





\vspace{6pt} 



\authorcontributions{Data Analysis and Writing—Original Draft Preparation, M.K. and K.Z.; Fabrication and Measurement of DNTT TFTs, J.B. and H.K.; Fabrication and Measurement of Pentacene and C60 TFTs, A.P., E.K-P., G.S and B.S.; Editing J.B., H.K., M.K. and K.Z.}

\funding{This research was funded by FWF grant number I 2081-N20.}


\conflictsofinterest{The authors declare no conflict of interest.}
\externalbibliography{yes}
\bibliography{Check_OTFT_model_A}



\end{document}




\begin{figure}[H]
\centering
\includegraphics[width=0.95\textwidth]{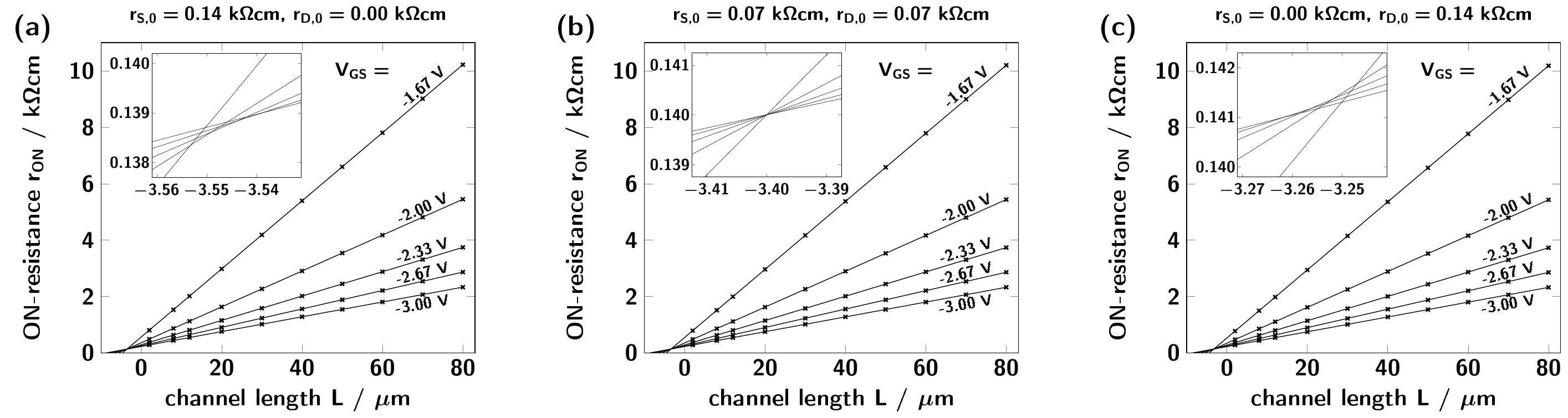}
\caption{Transmission line method (TLM)\cite{kanicki_performance_1991,luan_experimental_1992} performed on a simulated set of devices with the parameters extracted for the real set of devices with TLM (intrinsic channel mobility $\mu=3.2$~cm$^{2}$/Vs, threshold voltage $V_{T}=-1.25$~V, transfer length $L_{T}=3.4$~$\mu$m and combined contact resistance $r_{C,0}=r_{S,0}+r_{D,0}=0.14$~k$\Omega$cm) apportioning $r_{C,0}$ in three different ways to the source resistance $r_{S,0}$ and drain resistance $r_{D,0}$. The TLM is evaluated at a non-vanishing drain-source voltage $V_{DS}=-0.1$~V to visualize the generated error done by not taking $V_{DS}\rightarrow 0$~V. For each simulated device, the drain current $I_{D}$ is calculated at $V_{DS}=-0.1$~V and gate-source voltages $V_{GS}=$-1.67~V, -2.00~V, -2.33~V, -2.67~V and -3.00~V. Those drain currents are used to get an estimate for the ON-resistance $r_{ON}=WV_{DS}/I_{D}$. The such calculated values for $r_{ON}$ were used to perform a TLM. The inserts show a magnification of the intercept of all fit lines in the region of negative channel lengths $L$. In (\textbf{a}), the entire contact resistance $r_{C,0}$ was assigned to the source side, resulting in an intercept that is smeared out towards more negative channel lengths. The extracted parameters of $\mu=3.197$~cm$^{2}$/Vs, $V_{T}=-1.300$~V, $L_{T}=3.551$~$\mu$m and $r_{C,0}=0.1385$~k$\Omega$cm reflect this behaviour with a too high transfer length $L_{T}$ and a too low $r_{C,0}$. The threshold voltage is shifted by $V_{DS}/2$ and the mobility is nearly not affected. In (\textbf{b}), $r_{C,0}$ was equally distributed over $r_{S,0}$ and $r_{D,0}$, leading to a precise intercept. The extracted parameters $\mu=3.200$~cm$^{2}$/Vs, $V_{T}=-1.300$~V, $L_{T}=3.400$~$\mu$m and $r_{C,0}=0.14$~k$\Omega$cm perfectly match the input parameters except the threshold voltage which is shifted by $V_{DS}/2$. In (\textbf{c}), $r_{C,0}$ was entirely attributed to $r_{D,0}$, giving rise to an intercept smeared out towards more positive channel lengths. The extracted parameters are changed in the opposite direction compared to (\textbf{a}): $\mu=3.203$~cm$^{2}$/Vs, $V_{T}=-1.300$~V, $L_{T}=3.249$~$\mu$m and $r_{C,0}=0.1414$~k$\Omega$cm. Only the threshold voltage is shifted in the same way as in (\textbf{a}) and (\textbf{b}) by $V_{DS}/2$.}
\label{fig:SM_VDS_not_0}
\end{figure}

\begin{figure}[H]
\centering
\includegraphics[width=0.95\textwidth]{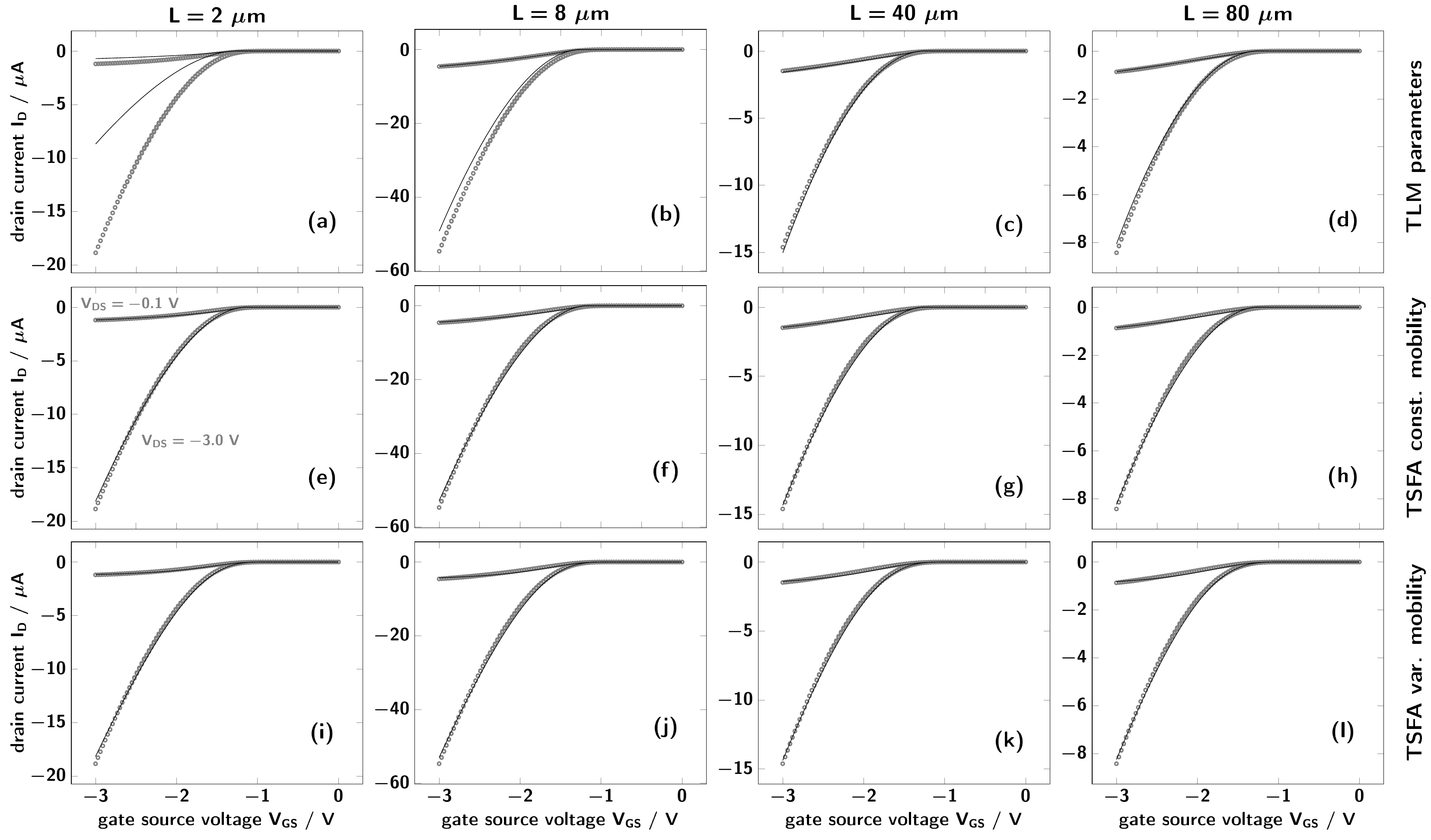}
\caption{Transfer characteristics corresponding to the output characteristics shown in Figure~3 and Figure~5 in the main manuscript. The measured values are drawn as gray symbols and the fitted characteristics are shown as black lines. The four different channel lengths $L$ refer to the shortest ($L=2$~$\mu$m), the second shortest ($L=8$~$\mu$m), an intermediate ($L=40$~$\mu$m) and the longest ($L=80$~$\mu$m) channel (from left to right). The topmost plots show the calculated characteristics corresponding to the parameters extracted by TLM, the middle ones to the TSFA with constant mobility and the lowermost ones to the TSFA with field- and charge-carrier-density-dependent mobility. Compared to the output characteristics (cf. Figure~3 and Figure~5 in the main manuscript), the deviations of the calculated curves from the measured ones are more difficult to see. Only for the top left characteristics (TLM parameters for $L=2$~$\mu$m), a disagreement is clearly obvious. To see the differences between the different fitting parameters (corresponding to different rows), a close look at the branching point at a gate-source voltage of about $V_{GS} = -1.5$~V has to be taken. At this point, a slight improvement of the agreement of measured and calculated characteristics can be seen when going from top to bottom.}
\label{fig:SM_TC_TLM_FA1_FA2}
\end{figure}

\begin{figure}[H]
\centering
\includegraphics[width=0.95\textwidth]{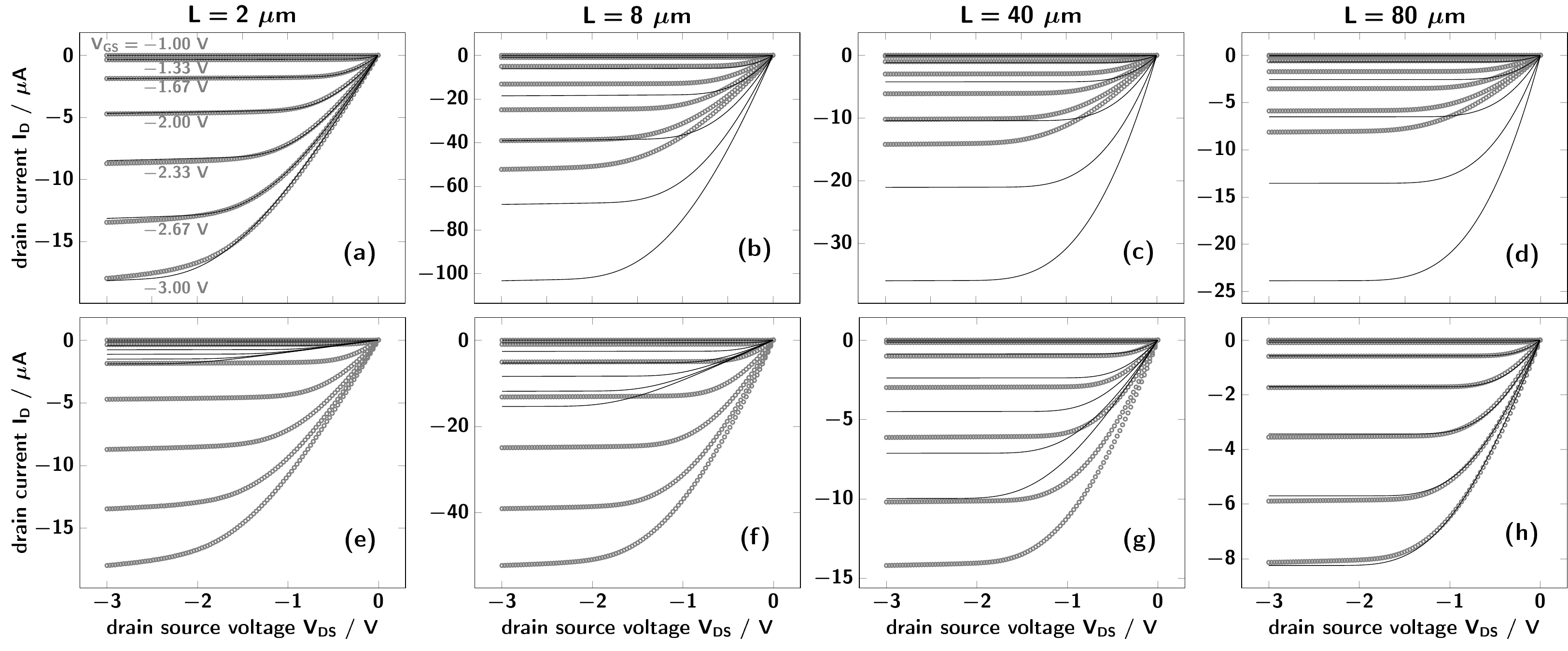}
\caption{Measured output characteristics of DNTT-based bottom-gate, bottom-contact thin-film transistors (TFTs) with the shortest ($L=2$~$\mu$m), the second shortest ($L=8$~$\mu$m), an intermediate ($L=40$~$\mu$m) and the longest ($L=80$~$\mu$m) channel length plotted as gray line and calculated output characteristics as black lines. The parameters for the calculated characteristics are the ones fitted with the field- and charge-carrier-density-dependent mobility model except the source and drain resistance $r_{S,0}$ and $r_{D,0}$. In the first row, the contact resistances of the TFT with the smallest channel length ($L=2$~$\mu$m) are used and in the second row, the contact resistances of the longest channel ($L=80$~$\mu$m) are used for all calculated characteristics. In (\textbf{a}) and (\textbf{h}), the contact resistances are the optimized ones, respectively, resulting in a match of measurement and calculation. From (\textbf{b}) to (\textbf{d}), the too low contact resistances cause an increasing overestimation of the calculated drain current $I_{D}$ and from (\textbf{e}) to (\textbf{g}), the drain current is underestimated due to too high contact resistances. Summarizing, the choice of the contact resistance has an important influence on the output characteristics underlining the fact that the channel-length dependence of the contact resistance is indeed significant.}
\label{fig:SM_OC_same_rC}
\end{figure}

\begin{figure}[H]
\centering
\includegraphics[width=0.75\textwidth]{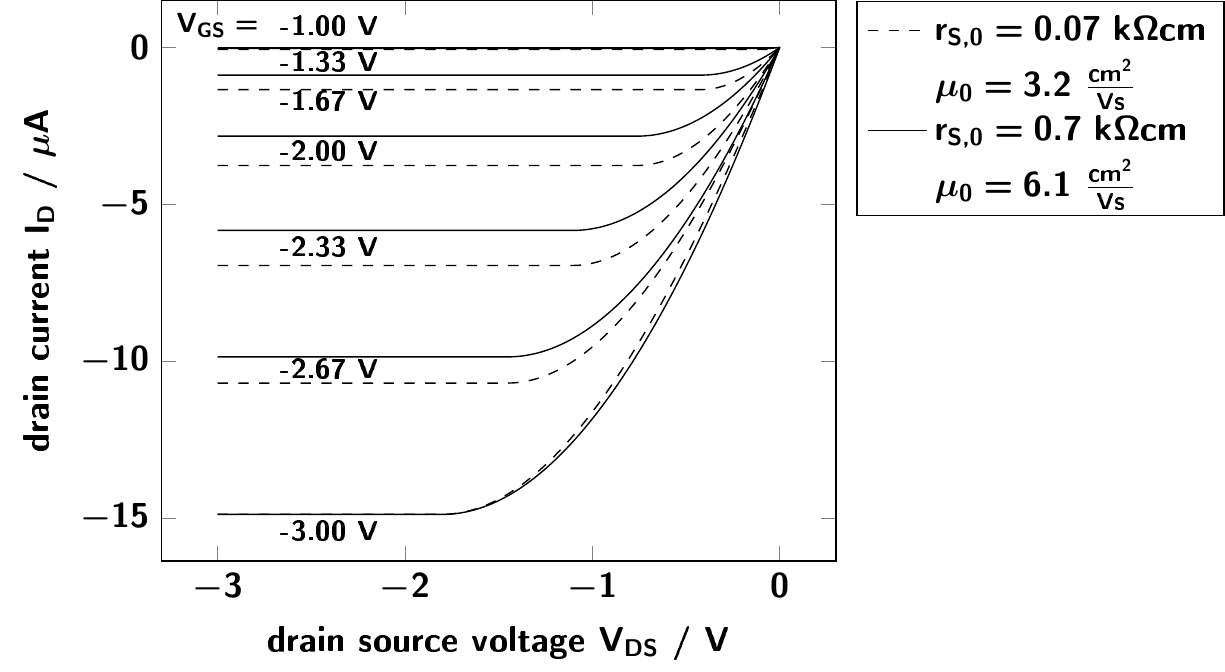}
\caption{Calculated output characteristics (solid line) for a TFT with channel length $L=40$~$\mu$m assuming the parameter set extracted by TLM (intrinsic channel mobility $\mu=3.2$~cm$^{2}$/Vs, threshold voltage $V_{T}=-1.25$~V, transfer length $L_{T}=3.4$~$\mu$m and source and drain resistance $r_{S,0}=r_{D,0}=0.07$~k$\Omega$cm). To show the correlation between contact resistance and mobility, a second output characteristics (dashed line) is plotted, for which only the mobility $\mu=6.1$~cm$^{2}$/Vs and the source resistance $r_{S,0}=0.7$~k$\Omega$cm are changed. Simultaneously increasing $\mu$ and $r_{S,0}$ results in a nearly perfect conformity of the output characteristics belonging to the highest gate-source voltage $V_{GS}=-3.00$~V and a pronounced shift of the saturation currents for all other $V_{GS}$. For a low contact resistance, the saturation current increases quadratically with $V_{GS}$ whereas for a high contact resistance, the saturation current increases linearly with $V_{GS}$. As a consequence, increasing $\mu$ and $r_{S,0}$ makes the spacing between $I_{D}$ in the saturation regime more uniform, which is able to partly compensate an incorrect mobility model.}
\label{fig:SM_V_G_spacing}
\end{figure}

\vspace{6pt} 

\externalbibliography{yes}
\bibliography{SM_check_OTFT_model_A}


